\documentclass[aps,prb,superscriptaddress,amsmath,amssymb,floatfix,twocolumn]{revtex4-1}
\pdfoutput=1
\usepackage{times}
\usepackage{bbold}
\usepackage{graphicx}
\usepackage{color}
\usepackage{hyperref}

\begin{document}

\title{Topological  density wave states of non-zero angular momentum}
\author{Chen-Hsuan Hsu}
\affiliation{Department of Physics and Astronomy, University of
California Los Angeles\\ Los Angeles, California 90095-1547}
\author{S. Raghu}
\affiliation{Department of Physics and Astronomy, Rice University, Houston, Texas 77005}
\author{Sudip Chakravarty}
\affiliation{Department of Physics and Astronomy, University of
California Los Angeles\\ Los Angeles, California 90095-1547}

\date{\today}

\begin{abstract}
The pseudogap state of high temperature superconductors is a profound mystery. It has tantalizing evidence of a number of broken symmetry states, not necessarily conventional charge and spin density waves. Here we explore a class of more exotic density wave states characterized by  topological properties observed in recently discovered topological insulators. We suggest that these rich topological density wave states deserve closer attention in not only high temperature superconductors but in other correlated electron states.
\end{abstract}

\pacs{}

\maketitle

\section{Introduction}
In a paper in 2000 Nayak~\cite{*[{We shall depart slightly from the  notation of }] Nayak:2000} provided an elegant classification of density wave states of non-zero angular momentum. The surprise is that given the roster of multitude of such states, so few are experimentally observed. Of these, the 
angular momentum $\ell=2$, spin-singlet has taken on a special significance in the context of pseudogaps in cuprate high temperature superconductors,~\cite{Chakravarty:2001}
It breaks translational symmetry, giving rise to a momentum dependent $d_{x^{2}-y^{2}}$ (DDW) gap, without
 modulating charge or spin, but  alternating  circulating charge currents from a plaquette to plaquette much like an antiferromagnet. In its pristine form,  in the half filled limit, that is,  for one electron per site, the Fermi surface of DDW consists of four Dirac points and is therefore  a semimetal. This broken symmetry state  has inspired much effort in characterizing the pseudogap  as a phase with an order parameter distinct from fluctuating superconducting order parameter. 
 
Presently, it appears from many experiments that the pseudogap may be susceptible to a host of possible competing orders.  Thus it is important and interesting to explore  an order parameter closely related to the singlet DDW, which retains many of its primary signatures such as the broken translational symmetry  or a particle-hole  condensate of higher angular momentum.  
In particular we consider  a density wave of non-zero angular momentum 
of  mixed singlet and triplet variety such that in the half-filled limit, it is a gapped insulator. 
Unlike the semimetallic DDW, it has   a non-vanishing
 quantized  spin Hall effect for a range of values of the chemical potential. This is in fact a topological Mott insulator~\cite{Raghu:2008} because it is the electron-electron interaction that is necessary for it to be realized. Further addition of charge carriers, doping,
 leads to  Lifshitz transitions destroying the quantization but not the very existence of the spin Hall effect. 
 
It is remarkable that such an unconventional broken symmetry, possibly relevant to  high temperature superconductors, belongs to the same class of currently discussed novel state of matter known as  topological insulators; in fact, our work is to some extent  motivated by these recent developments.~\cite{Hasan:2010,*Qi:2010} We wish to emphasize that the undoped parent compounds of high temperature superconductors are proven to be  antiferromagnets with sizeable moments and the spin density wave transforms according to $\ell=0$.~\cite{Chakravarty:1989} The proposed topological density wave should therefore be relevant at larger doping that perhaps originates from a nearby insulating state.  In no way is this different from the original suggestion of DDW.

\begin{figure}[htbp]
\begin{center}
\includegraphics[width=\linewidth]{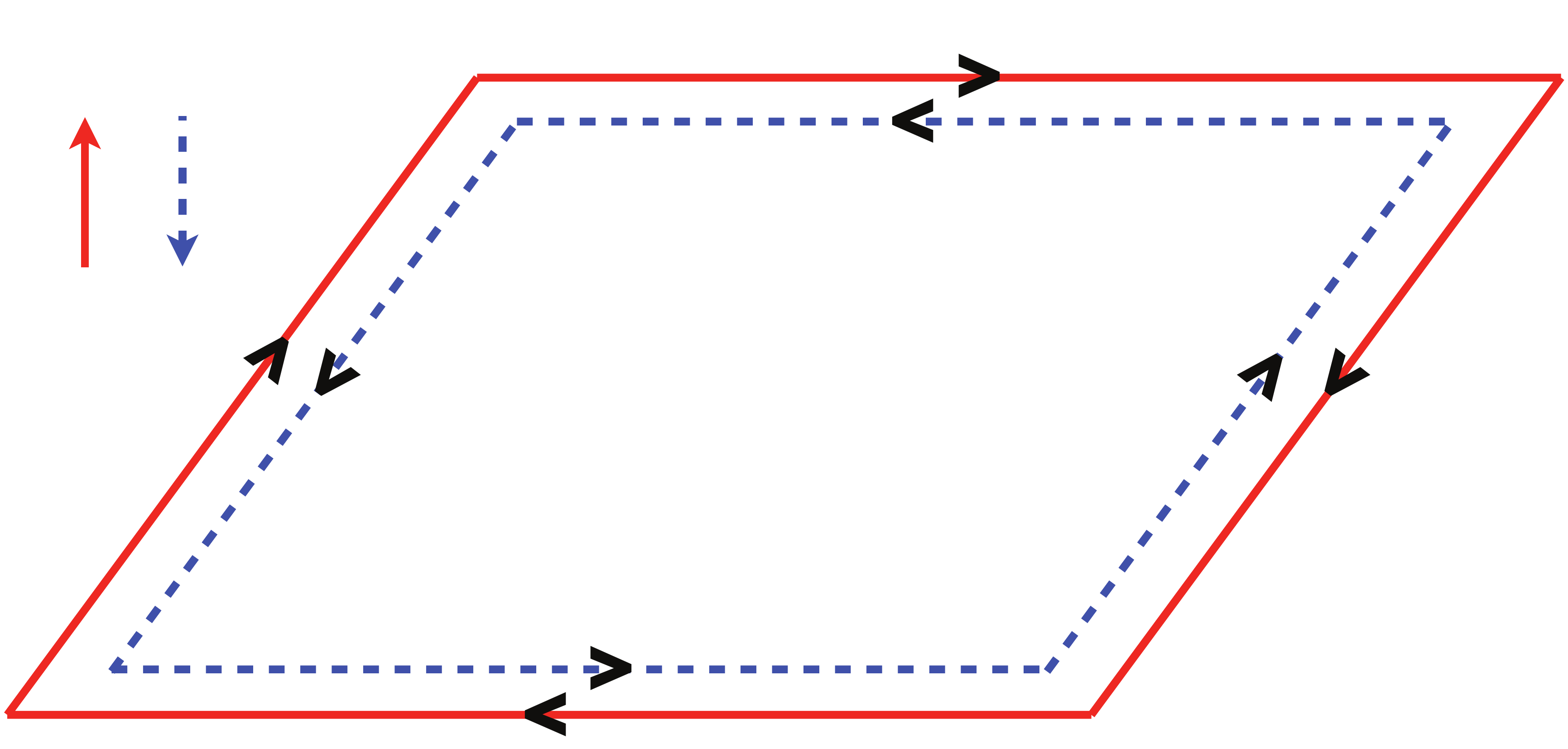}
\caption{(Color online) Triplet $i \sigma d_{x^{2}-y^{2}}$ density wave in the absence of an external magnetic field.  The current pattern of each spin species on an elementary plaquette is shown.   The state is a semimetal. On the other
hand $i\sigma d_{x^{2}-y^{2}}+ d_{xy}$ can be fully gapped for a range of chemical potential. An example is shown in Fig.~\ref{band}. }
\label{tDDW}
\end{center}
\end{figure}

It has been known that  triplet $i\sigma d_{x^{2}-y^{2}}$ order parameter corresponds to staggered circulating spin currents around a square plaquette.~\cite{Nersesyan:1991} wherein the oppositely aligned  spins circulate in opposite directions, as shown in Fig.~\ref{tDDW}. This reminds us of topological band insulators where oppositely aligned edge-spins travel in opposite directions. However, there is no topological protection because the bulk is not gapped, but is a semimetal instead. A more interesting  case is the order parameter $(i \sigma d_{x^{2}-y^{2}}+ d_{xy})$, where $\sigma=\pm 1$ for up and down spins, with the quantization axis along $\hat{z}$. Such a state not only satisfies time reversal invariance but is also fully gapped, analogous to time reversal invariant band insulators discovered recently.  
Singlet chiral $(i d_{x^{2}-y^{2}}+d_{xy})$ density wave  that breaks macroscopic time reversal symmetry was employed to deduce possible polar Kerr effect and anomalous Nernst effect~\cite{Tewari:2008, *Zhang:2009,*Kotetes:2008,*Kotetes:2010} in the pseudogap phase of the cuprates. Another topological state with a different symmetry of the order parameter was discussed in Ref.~\onlinecite{Ran:2008}

As to topological properties of superfluids,  we refer the reader to the book by Volovik.~\cite{Volovik:2003} Superconductors are particle-particle condensates, and,  as such, the orbital wave function constrains the spin wave function because of the exchange symmetry. What we are discussing here are particle-hole condensates, and there is no exchange requirement between a particle and a hole. Thus, orbital wave function {\em cannot} constrain the spin wave function. Thus an orbital singlet can come in both spin singlet and triplet varieties.

The plan of the paper is as follows: Section II is divided into three parts. Part A discusses  the topological aspects in the absence of magnetic field, while Part B contains results for a 
perpendicular magnetic field. The Part C consists of a thorough discussion of the bulk-edge correspondence that follows  from topological considerations. In section III we discuss Fermi surface reconstruction via a Lifshitz transition as the system is doped.  In section IV  possible
experimental detection schemes are suggested.  The symmetry of the order parameter that we have introduced is such that the necessary experimental  techniques  are more subtle than the detection
of more common broken symmetries, such as spin or charge density waves.

\section{Order parameter topology}

\subsection{Zero external magnetic field}

The order parameter that we consider is
\begin{equation}
\langle c_{{ k+Q},\sigma}^{\dagger} c_{{k},\sigma'}\rangle=(\Phi^{\mu}(k) \tau^{\mu})_{\sigma\sigma'},
\end{equation}
where $c_{{k},\sigma}^{\dagger}$($c_{{k},\sigma}$) is the Fermion creation(annihilation) operator with momentum ${k}$ and spin component $\sigma$; $\mu = 0,\cdots 3$,  $\tau_{1}$,  $\tau_{2}$, and $\tau_{3}$ are the standard Pauli matrices and $\tau_{0}=\mathbb{1}$. The nesting vector $\vec{Q} = (\pi/a,\pi/a)$.
We choose   the components of the order parameter  to be 
\begin{eqnarray}
\Phi^{3}(k)&\propto & {i \frac {W_0}{2}} (\cos {k_x} - \cos {k_y}) \equiv i W_k \\
\Phi^{0}(k)&\propto & {\Delta_0} \sin {k_x} \sin {k_y} \equiv \Delta_k.
\end{eqnarray}
and the remaining components are set to zero. The right hand side is written in terms of the gap parameters  and the conversion involves suitable coupling constants, which we do not need to specify in a non-selfconsistent Hartree-Fock theory. The lattice spacing {\it a} is set to unity.

\begin{figure}[htbp]
\begin{center}
\includegraphics[width=\linewidth]{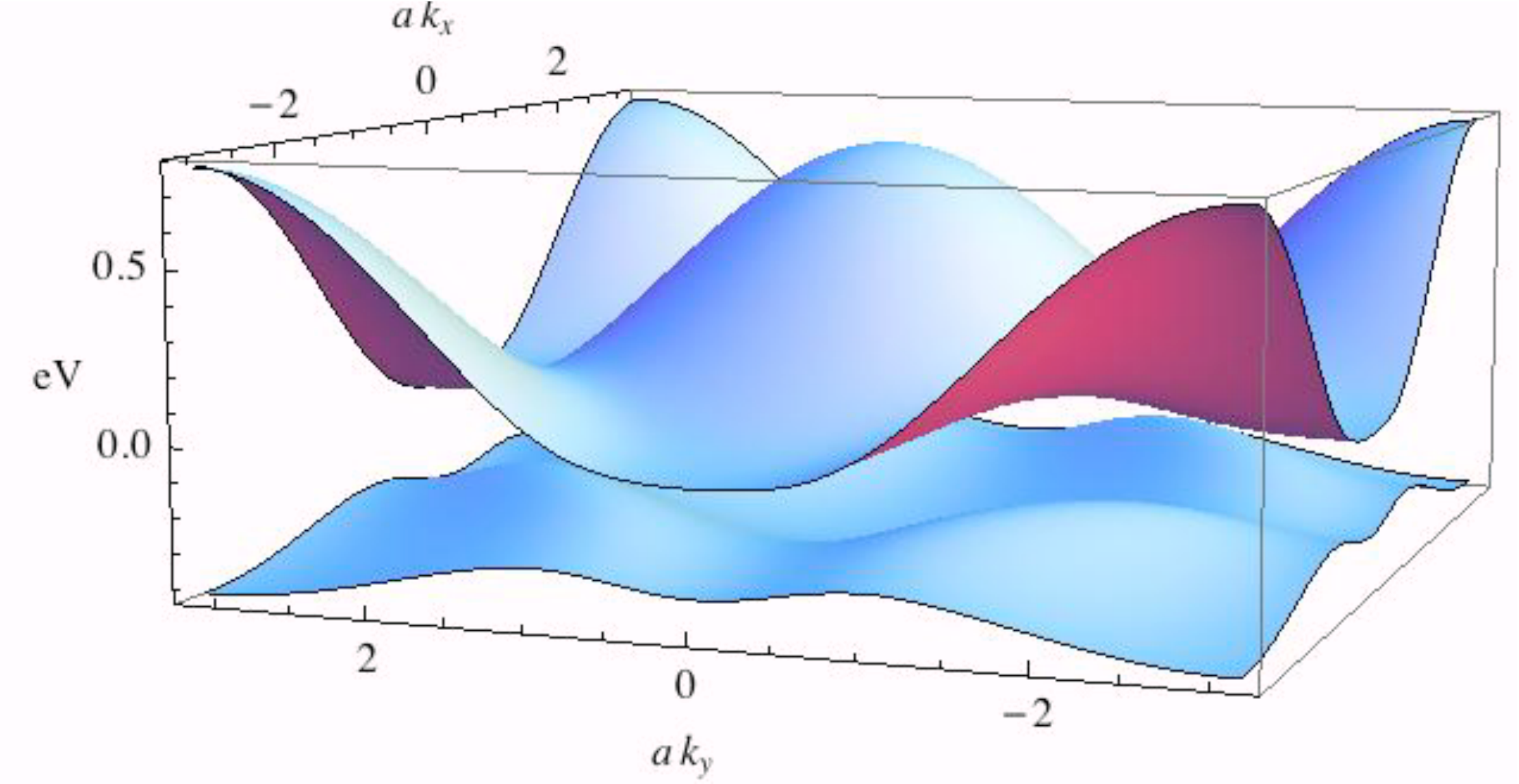}
\caption{(Color online) Energy spectra, $\lambda_{k,\pm}+\mu$,  corresponding  to $(i \sigma d_{x^{2}-y^{2}}+ d_{xy})$ density wave. Here, for illustration, we have chosen $W_{0}=t$ and $\Delta_{0}=-t$ and the band parameters, as described in the text. The chemical potential, $\mu$, anywhere within the spectral gap, the lower band is exactly a half-filled and the system is a Mott insulator, unlike the semimetallic DDW at half-filling.}
\label{band}
\end{center}
\end{figure}

In the absence of an  external magnetic field, the triplet $d\pm i d$ Hamiltonian is
\begin{equation}
\mathcal{H}_{d\pm i d}-\mu N =\sum_{k} \Psi_{k}^{\dagger} A_k \Psi_k,
\end{equation}
 where the summation is over the reduced Brilloin Zone (RBZ) bounded by $k_y \pm k_x = \pm\pi$, and the spinor, $ \Psi_{k}^{\dagger}$, is defined as $(c_{k,\uparrow}^{\dagger}, c_{k+Q,\uparrow}^{\dagger}, c_{k,\downarrow}^{\dagger}, c_{k+Q,\downarrow}^{\dagger}) $. The chemical potential is subtracted for convenience, $N$ being the number of particles.The matrix $A_k$ is
\begin{equation}
 A_{k} =\left(
  \begin{array}{cccc}
   \epsilon_{k}-\mu  & \Delta_k+iW_k      & 0                 & 0             \\
   \Delta_k-iW_k     & \epsilon_{k+Q}-\mu & 0                 & 0             \\
   0                 & 0                  & \epsilon_{k}-\mu  & \Delta_k-iW_k \\
   0                 & 0                  & \Delta_k+iW_k     & \epsilon_{k+Q}-\mu
  \end{array}
\right),
\end{equation}
with a generic set of band parameters, 
\begin{eqnarray}
\epsilon_k &=&\epsilon_{1k}+\epsilon_{2k}\\
\epsilon_{1k}&=&-2t(\cos k_x + \cos k_y), \; \epsilon_{2k}=4t^{\prime} \cos k_x \cos k_y.
\end{eqnarray}
We may choose $t=0.15\; eV$, renormalized by about a factor of 2 from band calculations and $t'=0.3 t$, and $W_{0}\sim -\Delta_{0}\sim t \sim J$, where $J$ is the antiferromagnetic exchange constant in high temperature superconductors, for the purpose of illustration.
Each of the two $2\times2$ blocks can be written in terms of two component spinors, $\psi_{k,\sigma}=(c_{k,\sigma}, c_{k+Q,\sigma})^{T}$, $\sigma=\pm1\equiv (\uparrow,\downarrow)$; for example, for the up spin block we have
\begin{equation}
\mathcal{H}_{\uparrow}=\sum_{k} \psi_{k,\uparrow}^{\dagger} \left[\mathbb{1}(\epsilon_{2k}-\mu)+\epsilon_{1k}\tau^{3}+\Delta_{k}\tau^{1}-W_{k}\tau^{2}\right]\psi_{k,\uparrow}
\end{equation}
The eigenvalues ($\pm$ refers to the upper and the lower bands respectively)
\begin{equation}
 \lambda_{k,\pm}= \epsilon_{2k}-\mu \pm E_{k}, \;
 E_{k}=\sqrt{\epsilon_{1k}^2 + W_k^2 + \Delta_k^2}.
\end{equation}
are plotted in Fig.~\ref{band}.
Since up and down spin components are decoupled, the Chern number for each component can be computed separately. 
%Note that while  $(\epsilon_{2k}-\mu)$ is present in the eigenvalues, it cannot enter the eigenvectors, because the identity operator commutes with the Pauli matrices.
After diagonalizing the Hamiltonian, we can obtain the eigenvectors
\begin{equation}
\Phi_{\sigma,\pm}({\bf k}) = (u_{\pm}{\it e}^{i \sigma \theta_k/2}, v_{\pm}{\it e}^{-i \sigma \theta_k/2})^T,
\end{equation}
 where
\begin{eqnarray}
u_{\pm}^2&=&{\frac{1}{2}}(1\pm{\frac{\epsilon_{1k}}{E_k}}), \\
v_{\pm}^2&=&{\frac{1}{2}}(1\mp{\frac{\epsilon_{1k}}{E_k}}), \\
\theta_k &=& \arctan({\frac{W_k}{\Delta_k}}) + \pi \Theta(-\Delta_k).
\end{eqnarray}

To compute the Berry phase of the eigenstates, we define the Berry curvature, $\vec{\Omega}_{\sigma,\pm}$ as
\begin{equation}
 \vec{\Omega}_{\sigma,\pm} \equiv i \vec{\bigtriangledown}_k \times \langle \Phi_{\sigma,\pm}^{\dagger}({\bf k})| \vec{\bigtriangledown}_k | \Phi_{\sigma,\pm}({\bf k}) \rangle
\end{equation}

Substituting the eigenstates into the above equation, the Berry curvature can be written as
\begin{eqnarray}
   \vec{\Omega}_{\sigma,\pm}&=& i \vec{\bigtriangledown}_k \times
   [ (u_{\pm}^2-v_{\pm}^2) \vec{\bigtriangledown}_k (i\sigma {\frac {\theta_k}{2}})].
\end{eqnarray}

Since $u_{\pm}$, $v_{\pm}$, and $\theta_k$ only depend on $k_x$ and $k_y$, only the z component, $\Omega_{\sigma,\pm}$, is non-zero, which is given by
\begin{eqnarray}
 \Omega_{\sigma,\pm} & = & \mp {\frac {\sigma}{2}} [ {\frac {\partial}{\partial k_x}}({\frac {\epsilon_{1k}}{E_k}})  {\frac {\partial \theta_k}{\partial k_y}} - {\frac {\partial}{\partial k_y}}({\frac {\epsilon_{1k}}{E_k}}) {\frac {\partial \theta_k}{\partial k_x}}] \nonumber \\
&=& \pm \sigma {\frac {1}{2E_k^3}}
 \left| \begin{array}{ccc}
  \Delta_k & W_k & \epsilon_{1k} \vspace{0.1in} \\
  {\frac {\partial{\Delta_k}}{\partial k_x}} & {\frac {\partial{W_k}}{\partial k_x}}&{\frac {\partial{\epsilon_{1k}}}{\partial k_x}} \vspace{0.1in} \\
  {\frac {\partial{\Delta_k}}{\partial k_y}} & {\frac {\partial{W_k}}{\partial k_y}}&{\frac {\partial{\epsilon_{1k}}}{\partial k_y}} \vspace{0.1in}
  \end{array}\right|.
\end{eqnarray}

From the above determinant, we can see that the Berry curvature will be zero if one of $\Delta_k$ and $W_k$ is zero, so we need a mixing of $d_{x^2-y^2}$ and $d_{xy}$ to have a non-trivial topological invariant. 

If we define the unit vector $\hat{n}_{\sigma} \equiv \vec{h}_{\sigma}/|\vec{h}_{\sigma}|$, where $\vec{h}_{\sigma} = (\Delta_k, -\sigma W_k, \epsilon_1)$, the Berry curvature can be written as
\begin{eqnarray}
 \Omega_{\sigma,\pm} &=& \mp {\frac {1}{2}} \hat{n}_{\sigma} \cdot ( {\frac {\partial{\hat{n}_{\sigma}}}{\partial k_x}} \times {\frac {\partial{\hat{n}_{\sigma}}}{\partial k_y}}).
\end{eqnarray}
More explicitly, the Chern numbers are
\begin{equation}
\begin{split}
N_{\sigma,\pm} &=  \int_{RBZ} {\frac{d^2\it{k}}{2 \pi }} \Omega_{\sigma,\pm}\\
& = \pm \sigma \int_{RBZ} {\frac{d^2\it{k}}{2 \pi }} {\frac {t W_0 \Delta_0} {E_k^3}} (\sin^2k_y + \sin^2k_x \cos^2k_y) \\
 &= \pm \sigma.
 \end{split}
\end{equation}
We can focus on the lower band as long as there is a gap between the upper and the lower bands. Then,
\begin{eqnarray}
  N & = &  N_{\uparrow,-} +  N_{\downarrow,-} = 0 \\
  N_{\text{spin}} & = & N_{\uparrow,-} -  N_{\downarrow,-} = (-1)-1 = -2
\end{eqnarray}
irrespective of the dimensionful parameters. Note, however, that the Chern numbers vanish unless both $\Delta_{0}$ and $W_{0}$ are non-vanishing. The quantization holds for a range of chemical potential $\mu$, as can be seen from Fig~\ref{band}. 

For the fully gapped case, there will be a quantized spin Hall conductance associated with the eigenstates. The ratio of the dimensions of the quantized spin Hall conductance to the quantized Hall conductance should be the same as the ratio of the  spin to the charge carried by a particle, since in two dimensions for both quantities the scale dependence $L^{d-2}$ cancels, that is,   
\begin{equation}
  \frac{[\sigma_{xy}^{\text{spin}}]}{[\sigma_{xy}]}  = \frac { {\frac{\hbar}{2}} }{e} .
\end{equation}
So, the quantized spin Hall conductance will be
\begin{eqnarray}
  \sigma_{xy}^{\text{spin}} = -  {\frac {e^2}{h}} {\frac {\hbar}{2e}} \, N_{\text{spin}} = {\frac {e}{2\pi}}    
\end{eqnarray}
The eigenstates, $|\Psi_{\sigma,\pm}(k) \rangle$, are also the eigenstates of ${\bf S}^2$ and ${\bf S_z}$ with eigenvalues $S^{2}={\frac{3}{4}}$ and  $S_z= - {\frac{\sigma}{2}}$. Since the spin $SU(2)$ is broken by the triplet DDW, one might wonder if the Goldstone modes not contained in the Hartree-Fock picture may not ruin the quantization. If $SU(2)$ is broken down to $U(1)$, then there is still a quantum number corresponding to, say $S_{z}$, which is transported by the edge currents in the system.  More succinctly, as long as time-reversal symmetry is preserved, we will still have Kramers degeneracy in our Hartree-Fock state, and therefore the edge modes will remain protected. 

\subsection{Non-zero magnetic field}
\begin{figure}[htbp]
\begin{center}
\includegraphics[width=\linewidth]{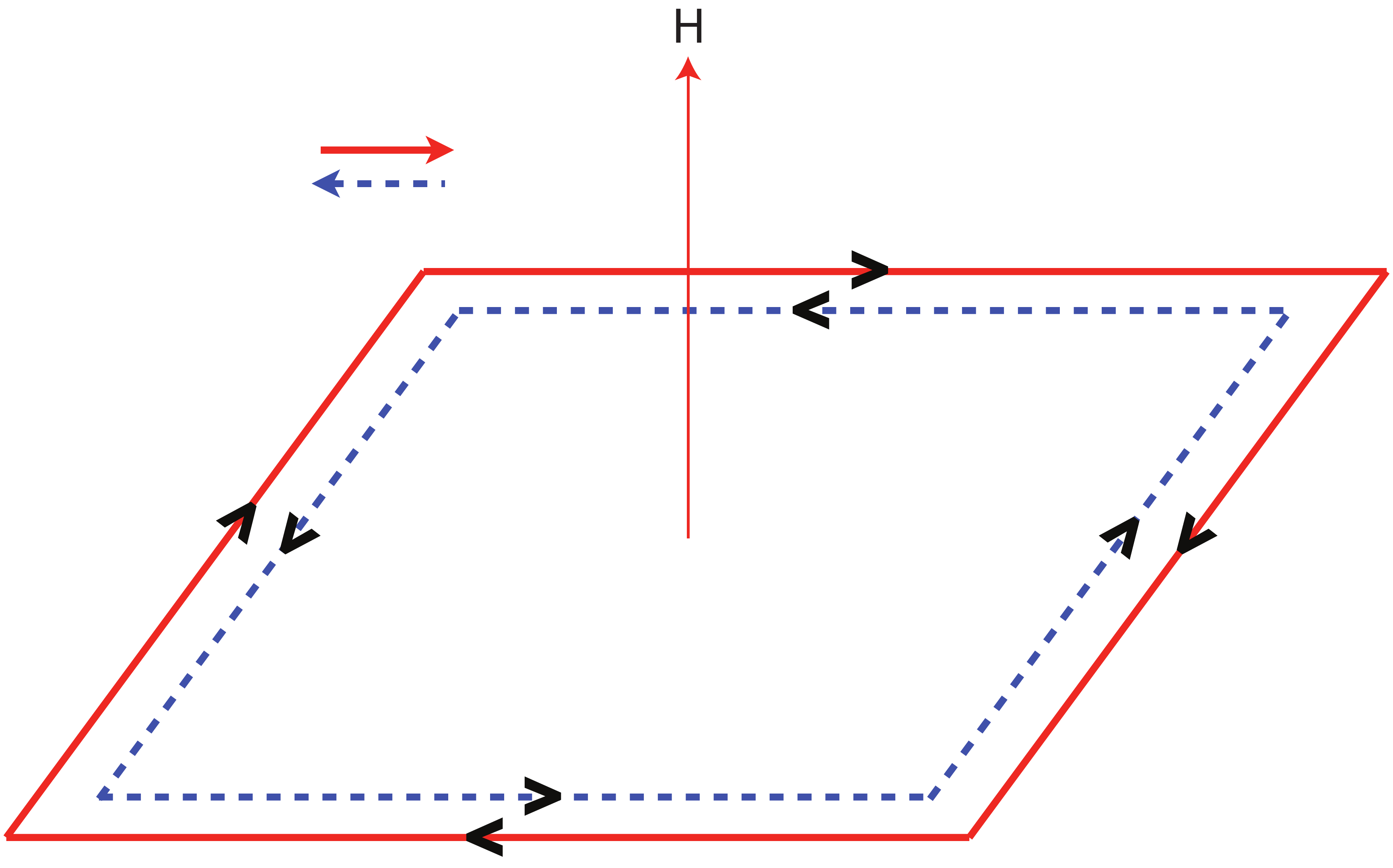}
\caption{(Color online) Spins are flopped perpendicular to the applied magnetic field H. Contrast with Fig.~\ref{tDDW}. }
\label{tripletddw-2}
\end{center}
\end{figure}

In an infinitesimal external magnetic field, $\vec{H}$, there will be a spin flop transition  in the absence of explicit  spin-orbit coupling, as shown in Fig.~\ref{tripletddw-2}. We can assume  $\vec{H} = H \hat{z}$ and the spins quantized along the $\hat{x}$ direction without any loss of generality. Then the Hamiltonian now becomes 
\begin{equation}
\mathcal{H}_{d\pm id}=\sum_{k} \Psi_{k}^{\dagger} A_k \Psi_k
\end{equation}
As before, the summation is over the RBZ, and the spinor is the same. The matrix $A_k$   is now \[
 A_k =\left(
  \begin{array}{cccc}
   \epsilon_{k,\uparrow}  & 0                       & 0                        & \Delta_k+iW_k \\
   0                      & \epsilon_{k+Q,\uparrow} &- \Delta_k-iW_k            & 0             \\
   0                      & -\Delta_k+iW_k           & \epsilon_{k,\downarrow}  & 0             \\
   \Delta_k-iW_k          & 0                       & 0                        & \epsilon_{k+Q,\downarrow}
  \end{array}
\right),
\]
where $\epsilon_{k,\sigma} = \epsilon_{k} + \sigma {\frac {g \mu_B H}{2}}=\epsilon_{k}+\sigma\gamma$. Although the spin up and down components are coupled, particles with momentum $k$ and spin up only couple to holes with momentum $k+Q$ and spin down, and vice versa. Therefore, by redefining the spinor, $ \Psi_{k}^{' \dagger} \equiv (c_{k,\uparrow}^{\dagger}, c_{k+Q,\downarrow}^{\dagger}, c_{k,\downarrow}^{\dagger}, c_{k+Q,\uparrow}^{\dagger}) $, the Hamiltonian can still be expressed as a block diagonal matrix: $\mathcal{H}_{d \pm id}=\sum_{k} \Psi_{k}^{' \dagger} A_k^{'} \Psi_k^{'}$. The Chern numbers for each subblocks, $i=1, 2$, can be calculated  as before. Therefore, defining  $\eta_i$ = + 1 or -1 for $i = 1$ or $2$, we obtain $E_{k,i} = [(\epsilon_1 + \eta_i \gamma)^2 + W_k^2 + \Delta_k^2 ]^{1/2}$, and the Berry curvature

\begin{eqnarray}
  \Omega_{i,\pm} & = & \mp {\frac {1}{2E_{k,i}^3}} \vec{h}_{i} \cdot ( {\frac {\partial{\vec{h}_{i}}}{\partial k_x}} \times {\frac {\partial{\vec{h}_{i}}}{\partial k_y}} ),
\end{eqnarray}
where $\vec{h}_{i} = (\eta_i \Delta_k, - W_k, \epsilon_1 + \eta_i \gamma)$. Performing a surface integration of the Berry curvature we get
\begin{equation}
\begin{split}
N_{i,\pm} &= \int_{RBZ} {\frac{d^2\it{k}}{2 \pi }} \Omega_{i,\pm} \\
          &= \pm \eta_{i} t W_0 \Delta_0 \int_{RBZ} {\frac{d^2\it{k}}{2 \pi E_{k,i}^3}}  \Big[ \sin^2k_y + \sin^2k_x \cos^2k_y  
          \\&- {\frac { \eta_i \gamma }{4t}} (\cos k_x \sin^2 k_y + \sin^2 k_x \cos k_y)  \Big]
 \end{split}
\end{equation}
The $\pm$ refers to the upper and the lower band respectively.
The integral does not depend on the external field, nor on magnitude of the parameters $t$, $W_0$, and $\Delta_0$. The Chern numbers are
\begin{equation}
\begin{split}
 N_{i,\pm} &= \pm \eta_{i},  \;  N_{total}=N_{1,-}+N_{2,-}=0,\\
 N_{spin}&=  N_{1,-} -  N_{2,-} = -2, 
 \end{split}
 \end{equation}
Once again the spin Hall conductance is quantized, but the charge quantum Hall effect vanishes. The flopped spins carry the same current as before. The corresponding spin Hall conductance, as long as the gap survives, is
\begin{equation}
\sigma_{xy}^{\text{spin}} =  \frac{e}{2\pi}.
\end{equation} 
The eigenstates, $|\Phi_{i,\pm}({k}) \rangle$, are the eigenstates of ${\bf S}^2$ with eigenvalues $S^{2}={\frac{3}{4}}$, but not eigenstates of ${\bf S_z}$ because of the mixing of up and down spins.  

\subsection{Bulk-edge correspondence}

For the  $(i \sigma d_{x^{2}-y^{2}}+ d_{xy})$ order the bulk-edge correspondence can be studied by open boundary condition in the $x$-direction but periodic boundary condition in the $y$-direction, that is, by cutting open the torus. The edge modes if they exist will reside on the ends of the cylinders. The cut then leads to a Hamiltonian
\begin{equation}
{\cal H} = \sum_{k_{y},i,j}\Psi_{i,k_{y}}^{\dagger}A_{ij}(k_{y})\Psi_{j,k_{y}},
\end{equation}
where the spinor is $\Psi_{i,k_{y}}=(c_{i,k_{y}\uparrow}c_{i,k_{y}+\pi \uparrow},c_{i,k_{y}\downarrow}c_{i,k_{y}+\pi \downarrow})^{T}$, and $A_{ij}(k_{y})$ is a $4N\times 4N$ matrix parametrized by the wave vector $k_{y}$, which is given by
\begin{eqnarray}
A_{ij}(k_y) &=& \left(
                 \begin{array}{cccc}
                 T_{ij}(k_y)                 & S_{ij,\uparrow}(k_y) & 0 & 0\\
                S_{ij,\uparrow}^{\dagger}(k_y) & T_{ij}(k_y+\pi)  & 0 & 0\\
                 0  & 0    & T_{ij}(k_y)     & S_{ij,\downarrow}(k_y)\\
                 0  & 0    & S_{ij,\downarrow}^{\dagger}(k_y) & T_{ij}(k_y+\pi)\\ 
                 \end{array}
              \right), \nonumber
\end{eqnarray}
where $T_{ij}(k_y)$ and $S_{ij,\sigma}(k_y)$ are $N \times N$ matrices:
\begin{widetext}
\begin{eqnarray} 
T_{ij}(k_y) &=& \left(
        \begin{array}{ccccc}
-\mu -2t \cos{k_y} & -t +2t' \cos{k_y}  & 0               & \cdots & \cdots  \\
-t +2t' \cos{k_y}  & -\mu -2t \cos{k_y} & -t +2t' \cos{k_y} & \cdots & \cdots  \\
0                & -t +2t' \cos{k_y}  & -\mu -2t \cos{k_y} & -t +2t' \cos{k_y}  & \cdots  \\
\vdots & \vdots & \vdots & \ddots          & -t +2t' \cos{k_y}    \\
       &        &        & -t +2t' \cos{k_y} & -\mu -2t \cos{k_y} 
         \end{array} 
         \right), \nonumber \\
S_{ij,\sigma}(k_y) &=& i \sigma \frac{W_0}{4} \left(
  \begin{array}{ccccc}
  -2 \cos{k_y} &   -1        &     0      & \cdots &  \\
     1       &   2 \cos{k_y} &     1      & \cdots &  \\
     0       &   -1        & -2 \cos{k_y} &    -1  & \cdots  \\
  \vdots     &  \vdots     & \vdots     & \ddots &  (-1)^{N-1}    \\
             &             &            & (-1)^N &(-1)^{N} 2 \cos{k_y} 
  \end{array} \right)
+ i \frac{\Delta_0}{2} \sin{k_y} \left(
  \begin{array}{ccccc}
       0  &  1  &    0   & \cdots &  \\
       1  &  0  &   -1   & \cdots &  \\
       0  & -1  &    0   &     1  & \cdots  \\
   \vdots & \vdots & \vdots & \ddots &  (-1)^{N}    \\
          &     &        & (-1)^N & 0
   \end{array} \right) \nonumber.
\end{eqnarray}
\end{widetext}
The corresponding one dimensional system with $N$ sites depends on the band structure and the order parameters defined above.

\begin{figure}[htbp]
\begin{center}
\includegraphics[width=\linewidth]{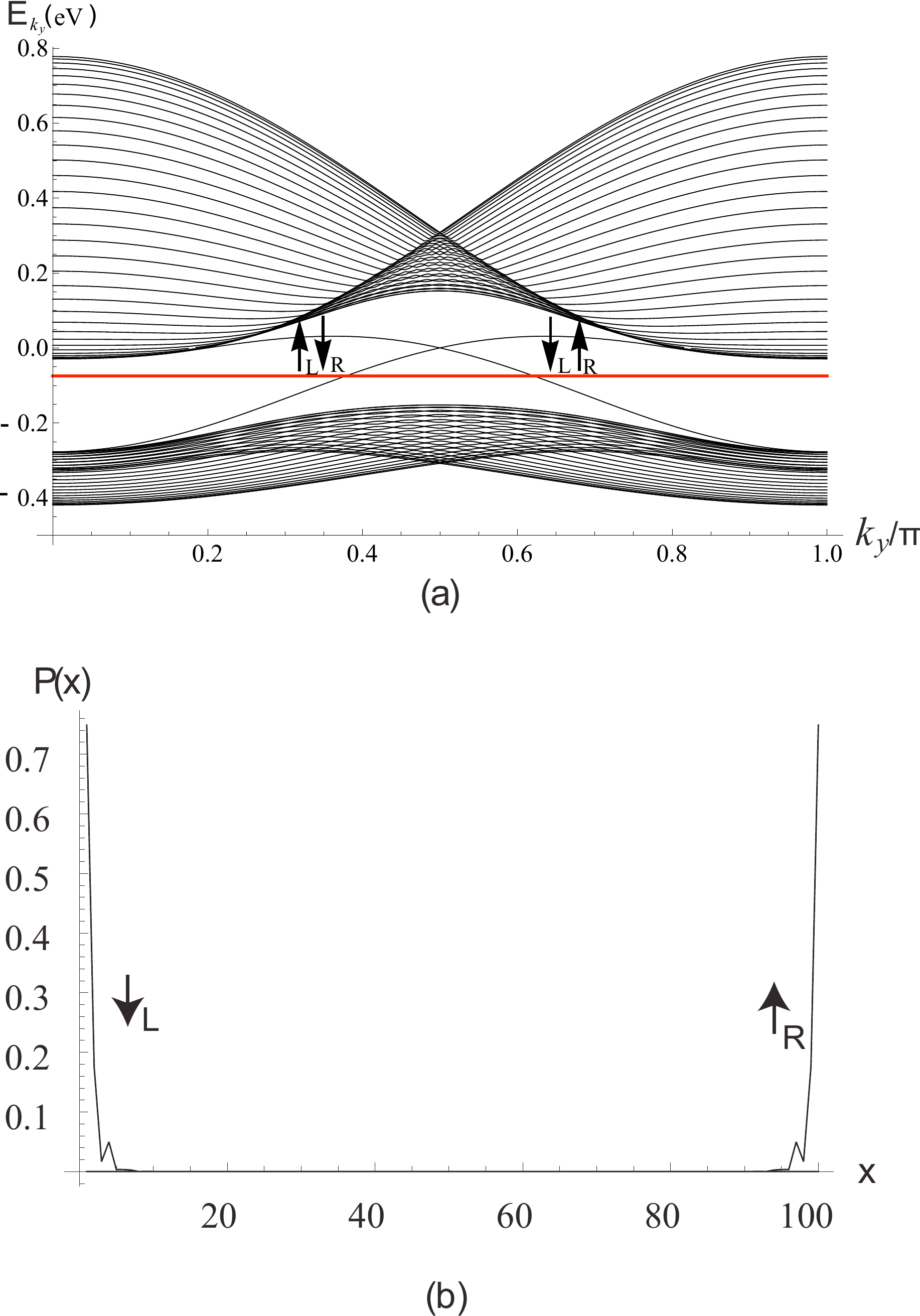}
\caption{(a) Spectrum of the triplet ($d\pm id$)-density wave on a cylinder.  Parameters are $t=0.15 eV, t'=0.3t, \mu = -0.075eV, W_0 = t$, and $\Delta_0 = -t$. The subscripts $L$ and $R$ to the spins correspond to left and right modes. (b) The probability density for positive group velocity for $L$ and $R$ spins for a lattice of $N=100$ sites.}
\label{edgestates}
\end{center}
\end{figure}

The eigenvalue spectra are shown in Fig~\ref{edgestates}. The spectra, degenerate for up and down spins, are plotted  in the range $0 \leq k_{y} \leq \pi$ ($k_{y}<0$ can be obtained by reflection).  To find the edge states we choose the chemical potential  in the gap. In Fig.~\ref{edgestates}, we put $\mu = -0.075eV$ for the purpose of illustration. There are two edge states with positive group velocity, one with up spin and the other with down spin.  Let them be $\psi_{>,\uparrow}$ and $\psi_{>,\downarrow}$, respectively. There are also two edge modes with negative group velocity denoted as $\psi_{<,\uparrow}$ and $\psi_{<,\downarrow}$ for up spin and down spin, respectively.  By explicitly computing the support of each of these wave functions, we have verified that electrons in states $\psi_{> , \downarrow}$ and $\psi_{<, \uparrow }$ are localized near the left edge of the system whereas those in states $\psi_{<, \downarrow}$ and $\psi_{>, \uparrow}$ are localized near the right edge. The localization length of these states is essentially a lattice spacing; an example is shown in Fig~\ref{edgestates}. 

\begin{figure}[htbp]
\begin{center}
\includegraphics[width=\linewidth]{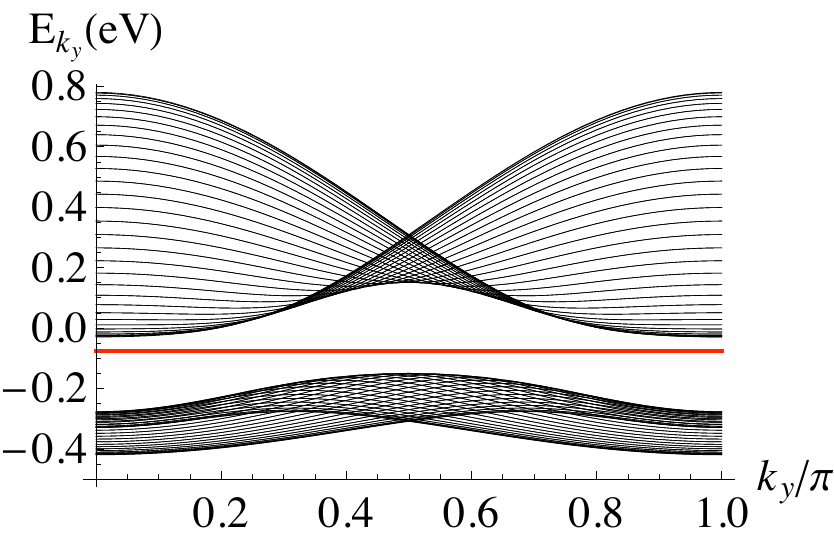}
\caption{The bulk spectra for fixed values of $k_{y}$ with the same parameters, as in Fig.~\ref{edgestates}.}
\label{bulkspectrum}
\end{center}
\end{figure}

It is interesting to see how this spectra compare with the one where periodic boundary conditions are applied in both $x$ and $y$ directions. After diagonalizing the Hamiltonian, we plot the spectra for a fixed value of $k_{y}$ for all values of the energies. The results are shown in Fig.~\ref{bulkspectrum}, which are essentially  identical to Fig.~\ref{edgestates}, except that the edge states are missing. 

\section{Fermi Pockets and Lifshitz transition}

\begin{figure}[htbp]
\begin{center}
\includegraphics[width=\linewidth]{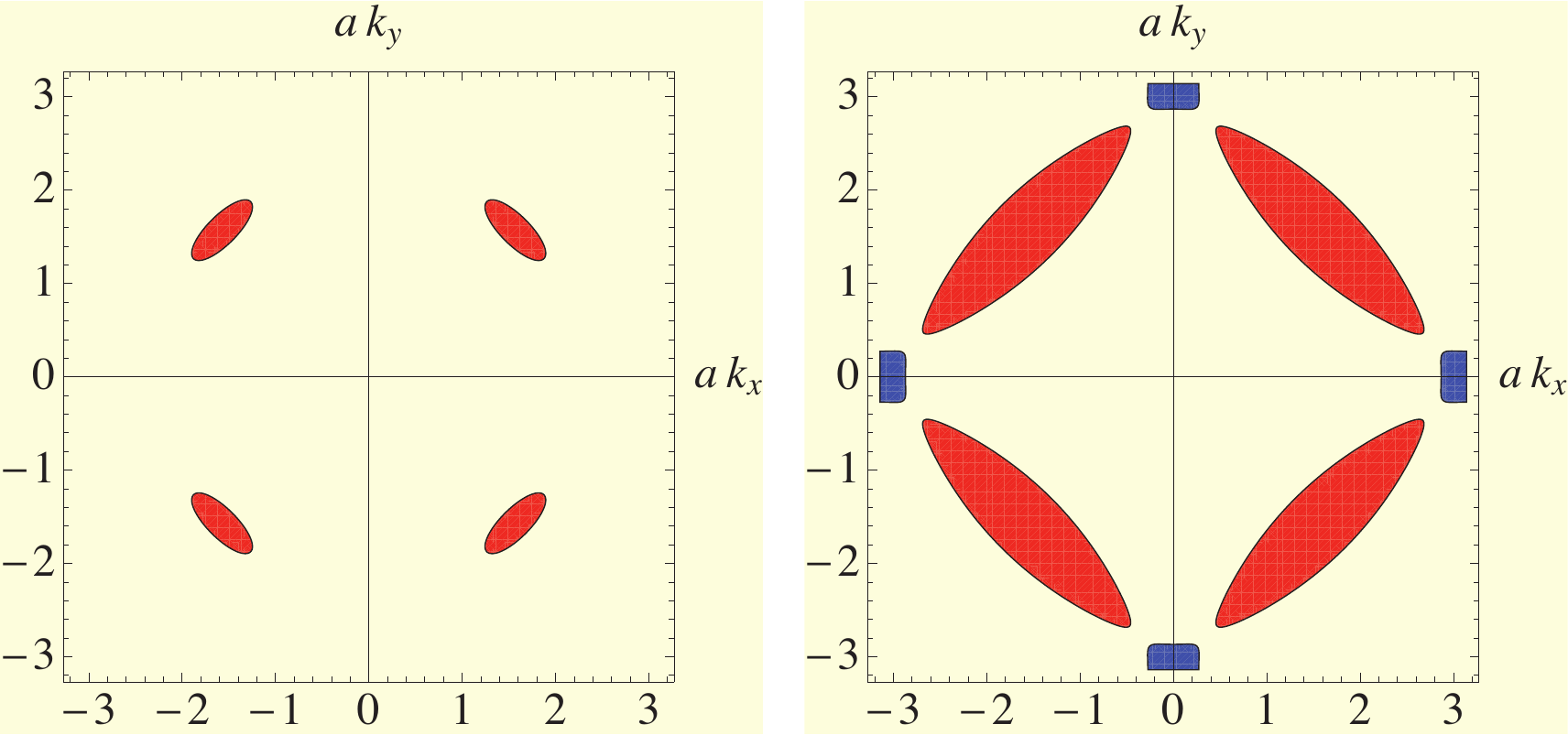}
\caption{(Color online) (Left) Region plot, $\mu=-0.16\;  eV$. Here for illustration, we have chosen $W_{0}=t$ and $\Delta_{0}=-t$ as before. The hole pockets open up. (Right) $W_{0}=0.05 t$ and $\Delta_{0}=-0.5 t$ illustrating the opening of the electron pockets at $(\pi,0)$ and symmetry related points with enlarged hole pockets.}
\label{electron-hole}
\end{center}
\end{figure}

It is interesting to track the evolution of successive Lifshitz transitions as we change the parameters. At first, when we lower the chemical potential, four hole pockets will open up in the full Brillouin zone, as shown in Fig.~\ref{electron-hole} and the corresponding spin Hall effect will lose its quantization but not the effect itself. But in mean field theory this cannot continue indefinitely with  the nodal or the antinodal gaps fixed. So the parameters $W_{0}$ and $\Delta_{0}$ will also decrease and will lead to a further opening of two electron pockets in the full Brillouin zone, as shown in Fig.~\ref{electron-hole}. Ultimately, when the doping is increased further, the large Fermi surface will emerge as a further Lifshitz transition. There is good evidence that such Lifshitz transitions indeed occur in high temperature superconductors.

\section{Experimental detection}

While there are many speculations about the nature of the pseudogap, they largely fall into two categories: 1) it is a crossover between a Mott insulator and a Fermi liquid, without any sharp, coherent excitations, and 2) it reflects a broken symmetry, with quasiparticles due to reconstructed Fermi surface that, despite strong correlations in the system, can behave in many ways as weakly interacting particles.  The resolution of this dichotomy will ultimately be settled by experiments, which, to date, have shown some support for both.  In the absence of a definitive evidence one way or the other, we have adopted the second perspective (to some extent motivated by recent quantum oscillation experiments) to see what consequences there may be of having a broken symmetry phase with sufficiently hidden order, in particular one that has striking similarities to topological insulators.  

A prime characteristic  of a broken symmetry is that deep in the broken symmetry phase, an effective mean-field, or a Hartree-Fock Hamiltonian,  suffices in discussing the properties of matter, and the symmetries alone determine the excitation spectra and the collective modes. It is only in the proximity of quantum critical points that such a description breaks down but that is not the subject of discussion here. Moreover, those properties that are determined by symmetries alone should be robust and can be understood in the weak coupling limit, simplifying our task of exploring correlated electron system.

The mixed triplet-singlet order parameters considered here is even more hidden than the corresponding singlet DDW. Not only do they not modulate charge or spin, but so long as spin-orbit coupling is absent, they are  also {\em invisible to elastic neutron scattering} because there is no associated staggered magnetic field, as in a singlet DDW. 

Inelastic neutron scattering can detect its signature in terms of a spin gap at low energies in the longitudinal susceptibility and signatures in the transverse susceptibility of quasi-Goldstone modes, and even onset of a finite frequency resonance mode. Recall that at any finite temperatures $SU(2)$ symmetry cannot be spontaneously  broken in two dimensions; interlayer coupling is necessary to stabilize it. Thus the scale of symmetry breaking must be considerably smaller than $t\sim J$, and the signature must be sought at higher energies. It could be a challenge to disentangle the signal from inelastic spin density wave excitations. On the other hand since the  quasiparticle excitations are essentially identical to the singlet DDW, the quantum oscillation properties will be  similar,~\cite{Chakravarty:2008b,*Doiron-Leyraud:2007} except perhaps those in a tilted field,~\cite{Garcia:2010,*Ramshaw:2011} which is currently being explored. The essence of this order parameters is modulation of spin current and kinetic energy. So, it will require probes that can detect higher order correlation functions, such as the two-magnon Raman scattering. In the presence of modest spin-orbit coupling, it may be possible to find small  shifts of nuclear quadrupolar frequency (NQR). The modulation of the kinetic energy arising from the $d_{xy}$ component, in particular staggered modulation of $t'$, may lead to anomalies in the propagation of ultrasound~\cite{[{A. Shekhter, personal communication; }] Bhattacharya:1988} at a temperature where such an order is formed, presumably at the pseudogap temperature $T^{*}$. The detection of the unique features of the proposed order parameter, the  spin Hall effect and edge currents would be even more challenging. 

The effects of non-magnetic impurities on the mixed triplet-singlet phase studied here are rather subtle.  We expect such disorder to couple only weakly to spin currents.   Generically, disorder will couple differently to the $i \sigma d_{x^2-y^2} $ and $d_{xy}$ components since each breaks a different symmetry.  However, by breaking both the point group and lattice translation symmetries, disorder can enable mixing with (generally incommensurate) density wave states in other angular momentum channels.  For example, at the level of Landau theory, we expect terms in the free energy proportional  to product of quadratic powers of the component order parameters, which would be proportional to the impurity concentration, thus inducing spin or charge density waves.  So long as spin rotational symmetry is preserved in the normal state, the phase transition into the $i \sigma d_{x^2-y^2}$ state can remain sharp.  

From the standpoint of topological order at zero temperature, the effects of weak disorder are somewhat simpler. Since the density wave phase considered here is a gapped phase with topological order that is protected by time-reversal symmetry, it remains robust against weak non-magnetic disorder.  Thus, the phase can still be described in terms of its topology at zero temperature, a feature which it shares with topological band insulators.  

Lastly, we remark that in the presence of magnetic impurities, the phase is not sharply defined - either as a broken symmetry or in terms of it's underlying topology.

In terms of microscopic models beyond the phenemenology discussed here, it is almost certain that correlated hopping processes will play a key role,~\cite{Nayak:2002} Finally, since $d_{x^2-y^2}$ and $d_{xy}$ are two distinct irreducible representations on a square lattice, generically they will each have their own transition temperatures, as dictated by Landau theory. The development of the $d_{xy}$ order parameter would be at a higher temperature compared to the  triplet component which breaks $SU(2)$ and therefore requires interlayer coupling. Thus it follows that when applied to cuprates there must be two transitions in the pseudogap regime. Since the topological phase studied here arises from spontaneous symmetry breaking, it can support charged skyrmion textures in analogy with.~\cite{Grover:2008} The properties of such textures and their transport signatures shall be the topic of a forthcoming publication.

\acknowledgments
This work is supported by NSF under the Grant  DMR-1004520. We thank Liang Fu, Pallab Goswami,  and Chetan Nayak for discussion.

\end{document}